\documentclass[oneside]{style/cernyrep}
\usepackage[utf8]{inputenc}
\DeclareUnicodeCharacter{2212}{-}

\usepackage{comment} 
\usepackage{amsmath}
\usepackage{color}

\usepackage{graphicx}
\usepackage{caption}
\usepackage{subfig}
\usepackage{subfloat}
\usepackage{multirow}
\usepackage{booktabs}
\usepackage{adjustbox}
\usepackage{rotating}
\usepackage{siunitx}
\DeclareSIUnit\clight{\text{c}}

\usepackage{nth} 

\usepackage[backend=biber,sorting=none, style=numeric-comp]{biblatex}

\usepackage{pdfpages}

\usepackage{epstopdf}
\usepackage{acronym}
\usepackage{balance}
\usepackage{authblk}
\usepackage{fancyhdr}
\usepackage{afterpage}
\usepackage{csvsimple}
\usepackage{pgfsys}

\usepackage{todonotes}

\usepackage{tcolorbox}

\usepackage{lineno}
\usepackage{placeins}

\usepackage{emptypage}
\usepackage{lipsum}

\usepackage{texnames}
\usepackage{ctable}
\usepackage[T1]{fontenc}

\usepackage{blindtext}

\usepackage[bookmarks, colorlinks=true, linktoc=page, pdftex, linkcolor=blue, citecolor=blue, urlcolor=blue]{hyperref}
\usepackage{float}
\usepackage{xcolor}
\usepackage[toc,page]{appendix}
\usepackage[bookmarks, colorlinks=true, pdftex, linkcolor=blue, citecolor=blue, urlcolor=blue]{hyperref}

\usepackage{times,lipsum}
\usepackage[margin=1in]{geometry}
\usepackage[onehalfspacing]{setspace}

\usepackage{arydshln} 
\usepackage{makecell}
\usepackage{listings}

\usepackage[hang,flushmargin]{footmisc}
\fancypagestyle{plain}{}
\newlength{\oddmarginwidth}
\newlength{\evenmarginwidth}
\fancyhfoffset[LO,RE]{\oddmarginwidth}
\fancyhfoffset[LE,RO]{\evenmarginwidth}
\bibliography{bib/mergeTDRbib.bib} 
\pdfoutput=1
\RequirePackage{latex/atlaslatexpath}

\graphicspath{{./logos/}{./figures/}{./figures_chapt2/}}

\usepackage{luxetdr-defs}
\usepackage{luxepixel-defs}

\PassOptionsToPackage{percent}{overpic}

\title{
\vspace*{-1.5cm}
\noindent
\makebox[\textwidth]{\includegraphics*[trim={0 19cm 0 0},clip]{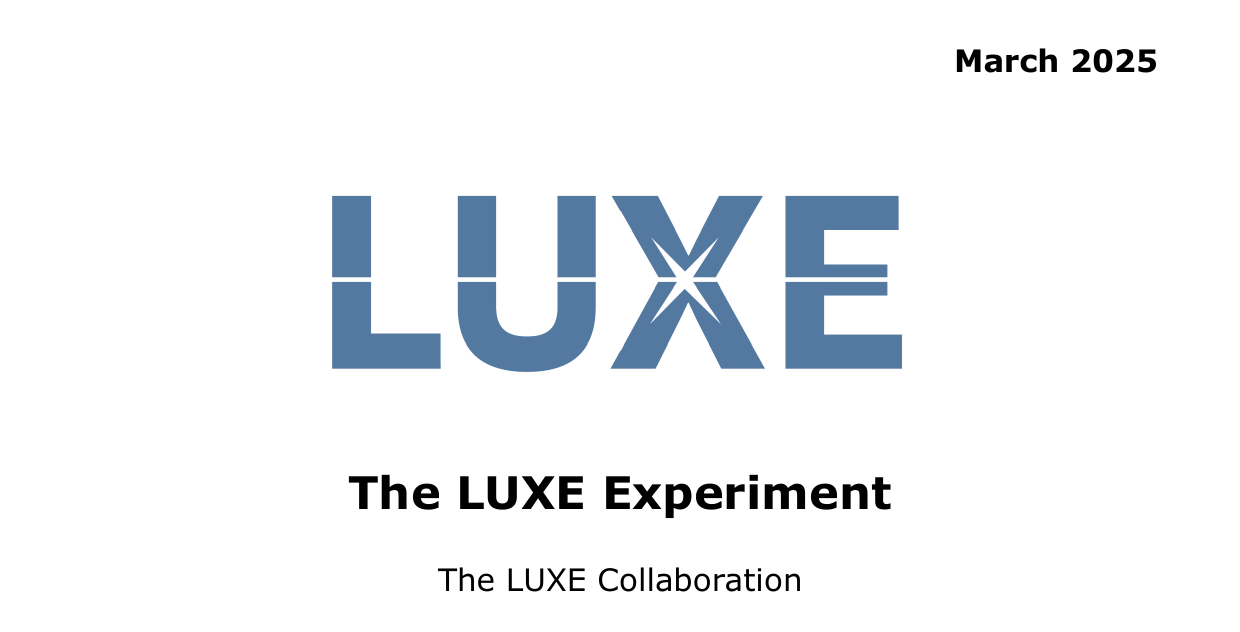}}
\vspace*{1.5cm}
}
\author{Contact persons: \\ 
\hspace{1cm} Matthew Wing\footnote{University College London, London, UK} (Spokesperson, m.wing@ucl.ac.uk) \\ 
\hspace{1cm} Ruth Jacobs\footnote{Deutsches Elektronen-Synchrotron DESY, Hamburg, Germany} (Editor, ruth.magdalena.jacobs@desy.de) \\
\hspace{1cm} Federico Meloni$^{\dagger}$ (Editor, federico.meloni@desy.de)
}
\begin{abstract}
    This document presents an overview of LUXE (Laser Und XFEL Experiment),  an experiment that will combine the high-quality and high-energy electron beam of the European XFEL with a high-intensity laser, to explore the uncharted terrain of strong-field quantum electrodynamics. The scientific case, facility, and detector setup are presented together with an overview of the foreseen timeline and expected capital costs.
\end{abstract}
\hypersetup{pdftitle={LUXE ESPPU 2025}, pdfauthor={The LUXE Collaboration}}
\begin{document}

\maketitle



\clearpage
\setcounter{page}{1}
\section*{Introduction}

The Laser Und XFEL Experiment (LUXE) is an experiment dedicated to measuring the products of collisions between ultra-relativistic electrons extracted from the European XFEL (EuXFEL) (or the high-energy $\gamma$-rays they generate in a target) and short, high-power laser pulses.
The scientific ambitions of LUXE are to break new ground in the exploration of Strong-Field Quantum Electrodynamics (SFQED) and to contribute to the search for particles and phenomena beyond the Standard Model of particle physics.

A set of detectors is designed to measure the properties of the electrons, positrons, and photons produced in the collisions of many-GeV electrons (or photons) with optical laser pulses at intensities where the charge-field interaction becomes non-perturbative. This will enable LUXE to make precision measurements of the transition from the perturbative to the non-perturbative regime of quantum electrodynamics.

The implementation of this project greatly expands the scope of the E144 experiment, performed at SLAC in the 1990's~\cite{Bamber:1999zt}, leveraging on the huge progress of high-power laser and accelerator technologies in the last three decades.

\section*{The scientific case}

QED has been tested to a very high precision in systems with weak electromagnetic (EM) fields. On the other hand, in EM fields of a strength comparable to or in excess of the scale identified by the ``Schwinger limit'' $\mathcal{E}_{\textrm{cr}}={m_{e}}^{2}c^{3}/(e\hbar)=1.32\times 10^{18}\,\textrm{Vm}^{-1}$, QED predicts many phenomena that have not yet been investigated experimentally~\cite{Fedotov:2022ely}. For example, in astrophysics, pair creation can accompany the gravitational collapse of black holes \cite{Ruffini:2009hg} and affect the propagation of cosmic rays \cite{nikishov62}. Some neutron stars are so strongly magnetised, that their magnetospheres probe the Schwinger limit \cite{Kouveliotou:1998ze,Harding_2006,Turolla:2015mwa}. In particle physics, in beam-beam collisions at future high-energy lepton colliders, strong-field effects are expected to feature prominently~\cite{yakimenko2019prospect,Bucksbaum:2020}, and probing of the Coulomb field of heavy ions can be sensitive to strong-field effects~\cite{Akhmadaliev:2001ik}. Strong fields, albeit in non-relativistic systems where the field strength scale is set by the ionisation potential, are also the target of much investigation in the atomic and molecular physics communities (see e.g. \cite{ivanov,Morgner:2023ghx}).

The Schwinger limit is currently orders of magnitude above any terrestrially producible field strength. However, in the rest frame of a high-energy probe particle, the EM field strength $\mathcal{E}$ is boosted by the Lorentz factor, $\gamma$, to $\mathcal{E}_{\ast}=\gamma \mathcal{E}(1+\cos\theta) \simeq 2\,\gamma \mathcal{E} $ if the collision angle $\theta$ is small (at LUXE $\theta=17.2^\circ$). In this way, by colliding a $16.5\,\textrm{GeV}$ EuXFEL electron beam (with Lorentz $\gamma$-factor $\simeq 3 \times 10^{4}$) with intense photon pulses produced by a laser, fields in the electron rest frame at LUXE can reach and exceed the Schwinger limit and hence SFQED phenomena can be accessed.

What sets SFQED phenomena apart from usual QED is described by two parameters, the ``classical nonlinearity parameter'' (or intensity parameter) $\xi$ and the ``quantum nonlinearity parameter'' (or strong-field parameter) $\chi$. Their ratio $\eta = \chi / \xi$ measures the energy of the collision between the probe particle and the laser field, also called the background field. In a plane-wave EM background, which approximates well the situation in a laser pulse as will be tested by LUXE, the electron-laser coupling can be described with $\xi=|e|\mathcal{E}\lambdabar_{e}/\hbar \omega_{L}$: the work done by the laser EM field over a reduced Compton wavelength of the electron in units of the laser photon energy. In weak fields, probabilities of QED events involving $n$ photons scale as $\sim \xi^{2n}$, with higher-order interactions being suppressed. When $\xi \sim O(1)$, this perturbative hierarchy breaks down. The $\chi$ parameter, which for an incident electron can be written as $\chi_{e} = \mathcal{E}_{\ast}/\mathcal{E}_{\textrm{cr}}$, is the ratio of the laser EM field, in the rest frame of the electron, to the Schwinger limit.  

The key processes and quantities to be addressed to characterise the SFQED regime are:

\begin{itemize}
\item The nonlinear Compton process. In this process, an electron absorbs a net number $n$ of laser photons, $\gamma_L$, from the laser background and converts them into a single, high-energy gamma photon $\gamma_{C}$:
\[
e^\pm +n\gamma_{L} \to e^{\pm} + \gamma_{C}.
\]

The magnitude of nonlinear and quantum effects can be clearly seen in the position of the `Compton edge'~\cite{Harvey:2009ry} in electron and photon spectra that varies with the intensity parameter, and where $\chi$ quantifies the electron recoil when it emits a photon. The measurement of the position of the Compton edge allows quantum and nonlinear effects to be to differentiated.

\item The nonlinear Breit-Wheeler process. In this process, an incoming photon absorbs the net number $n$ optical photons, $\gamma_L$, from the laser background and produces an $e^{+}e^{-}$ pair:
\[
\gamma + n\gamma_{L} \to e^{+} + e^{-}.
\]
Although linear Breit-Wheeler pair production was recently observed in Coulomb fields in heavy ion collisions \cite{STAR:2019wlg}, LUXE would provide the first observation of the nonlinear process in the non-perturbative regime, using real (as opposed to \textit{virtual}) photons.

\item The nonlinear trident process.  The nonlinear trident process combines a nonlinear Compton step and a nonlinear Breit-Wheeler pair-creation step off the radiated photon:
\[
e^{-} + n \gamma_{L} \to e^{-} + \gamma \qquad \textrm{and} \qquad \gamma + n'\gamma_{L} \to e^{+} + e^{-},
\]
where the number of background laser photons in the Breit-Wheeler step, $n'$, will in general be different to the net number of photons involved in the Compton step, $n$. 
Alternatively, this process can also take place as a one-step coherent production:
\[
e^{-} + n \gamma_{L} \to e^{-} + e^{+} + e^{-}.
\]
The relative abundance of the two diagrams depends on the length of the laser pulse, with the latter production mode being dominant in the case of short laser pulses.  Although observed in the perturbative multi-photon regime in the E144 experiment~\cite{Burke:1997ew}, and recently by NA63 in crystals~\cite{nielsen2022precision}, LUXE would be the first to measure nonlinear trident in the high-intensity non-perturbative regime using a source of real photons.
\end{itemize}

\begin{figure}[ht]
   \centering
   \includegraphics*[height=5.2cm]{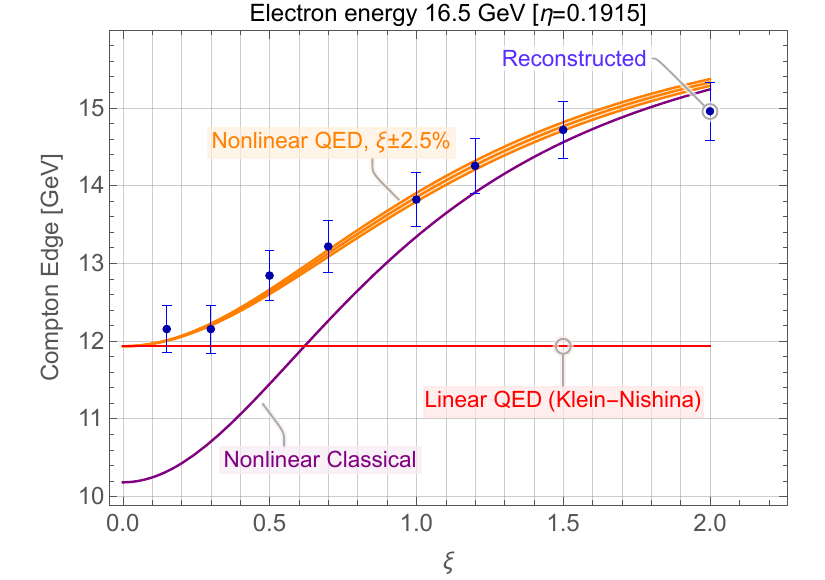}
   \includegraphics*[trim={0 6.35cm 0 0},height=5.25cm]{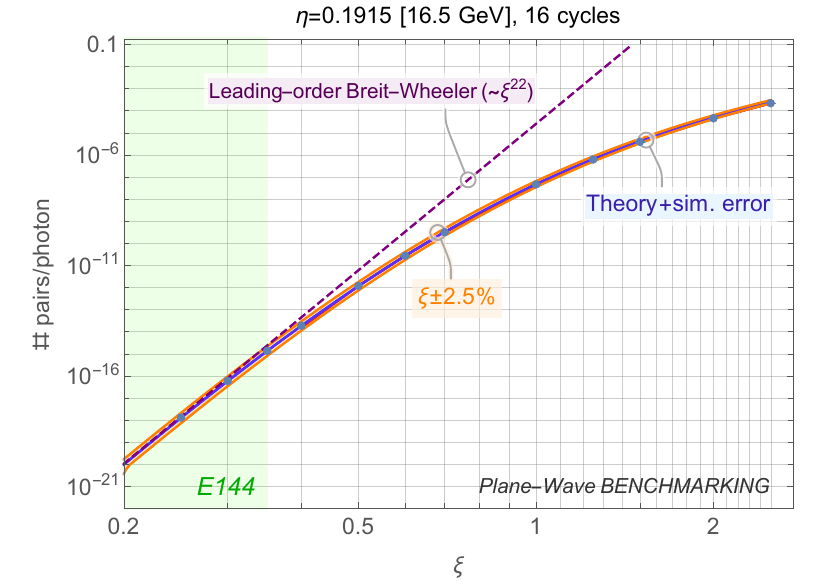}
   \caption{Left: expected experimental results for the measurement of the first Compton edge in the electron energy spectrum in \elaser interactions. Right: number of positrons produced in \glaser collisions as a function of $\xi$ as expected in the full QED calculation (solid line) and in the purely perturbative calculation (dashed line, in which the probability of pair creation scales as $\sim \xi^{2n}$). A 2.5\% uncertainty on $\xi$, corresponding to a 5\% uncertainty on the laser intensity, is illustrated in orange as uncertainty on the theoretical nonlinear QED predictions.}
   \label{fig:core_measurements}
\end{figure}

The experimental realm that LUXE intends to probe is unique in SFQED, because detailed quantitative theory predictions exist that can be confronted with experimental data. Figure~\ref{fig:core_measurements} shows the expected results for two key measurements: the position of the Compton edge in \elaser interactions and the multiplicity of $e^{+}e^{-}$ pairs per incoming photon in \glaser interactions. A new custom-made numerical simulation code called \textsc{Ptarmigan} \cite{ptarmigan} was written, in the absence of any other code being able to simulate physics in the full range of $\xi$, from $\xi \ll 1$ to $\xi \sim$ O(10). It does this by using the Locally Monochromatic Approximation (LMA)~\cite{heinzl.arxiv.2020}. The thorough benchmarking with more exact theory calculations from  plane-wave QED was essential to test the domain of applicability of this approximation. Although a few other codes exist that use the LMA (without benchmarking to the full QED result), they do not span this intensity range. For example, CAIN can only implement LMA simulations up to $\xi \sim$ 3, which is not sufficient for LUXE, since we plan to employ 40 to 350~TW lasers and reach $\xi \gtrsim$ O(10)~\cite{Blackburn:2021rqm,Blackburn:2021cuq}.

In addition to the precision SFQED programme, due to the high photon flux the LUXE experimental setup offers further opportunities to search for new phenomena beyond the Standard Model. There is significant experimental evidence for new physics beyond the Standard Model~\cite{McCullough:2018knz,Strategy:2019vxc}. Possible explanation of these experimental signals includes the existence of new, light, degrees of freedom, which are weakly coupled to the Standard Model and potentially long-lived. Axion-like particles (ALPs), which are generalisations of the posited axion that solves the strong-CP problem~\cite{Peccei:1977hh, Peccei:1977ur, Wilczek:1977pj, Weinberg:1977ma}, can couple to two photons and hence produce a signal in the LUXE experiment. A ``secondary production'' mechanism will be used, which involves using high-energy photons generated via the nonlinear Compton process, propagating downstream to a beam dump, in which ALPs can be created (via the Primakoff effect) and then decay on the other side of the dump to two photons.

A detector is then placed behind the dump to detect the signal photons while rejecting other beam-induced backgrounds due to e.g. neutrons and to measure the energies, positions, and angles of the photons. This aspect of LUXE is called LUXE-NPOD (LUXE New Physics search with Optical Dump) and described in more detail in Ref.~\cite{LUXEBSM}. 

LUXE-NPOD will be sensitive to the ALP-photon couplings in the region of $10^{-5}\,\textrm{GeV}^{-1}$ for pseudoscalar masses around $\sim 200\,\textrm{MeV}$, a parameter space as yet unexplored by running experiments. Using a photon beam also presents a novel and complementary way compared to classic dump experiments, e.g. compared to dumping the electron beam directly this method results in significantly lower backgrounds. 

\section*{European XFEL accelerator}
A very high-quality electron beam with a design energy of up to $\sim 17.5$~GeV is delivered by the linear accelerator of the European XFEL. The European XFEL has been operating according to specifications since 2017. The beam consists of $600 \, \mu s\,$-long bunch trains, each containing up to 2700 individual bunches, and the rate of bunch trains is normally 10~Hz. Only one of the bunches within a train will be used for the LUXE experiment as the laser operates with a frequency of 1~Hz (the remaining rate of non-colliding bunches will be used to measure beam-induced backgrounds in-situ). The removal of one bunch per bunch train is completely transparent to the experiments using the x-rays produced in the undulators.  For the purpose of this document, electron beam energies of 16.5 and 14.0~GeV are assumed. For LUXE, the highest beam energy is of most interest but it is also interesting to perform the measurement at several energies. The accelerator is designed for a bunch charge of 1~nC but mostly operates at 0.25~nC, corresponding to $1.5\times 10^9$ electrons. 

\begin{figure}[ht]
   \centering
   \includegraphics*[height=4.2cm]{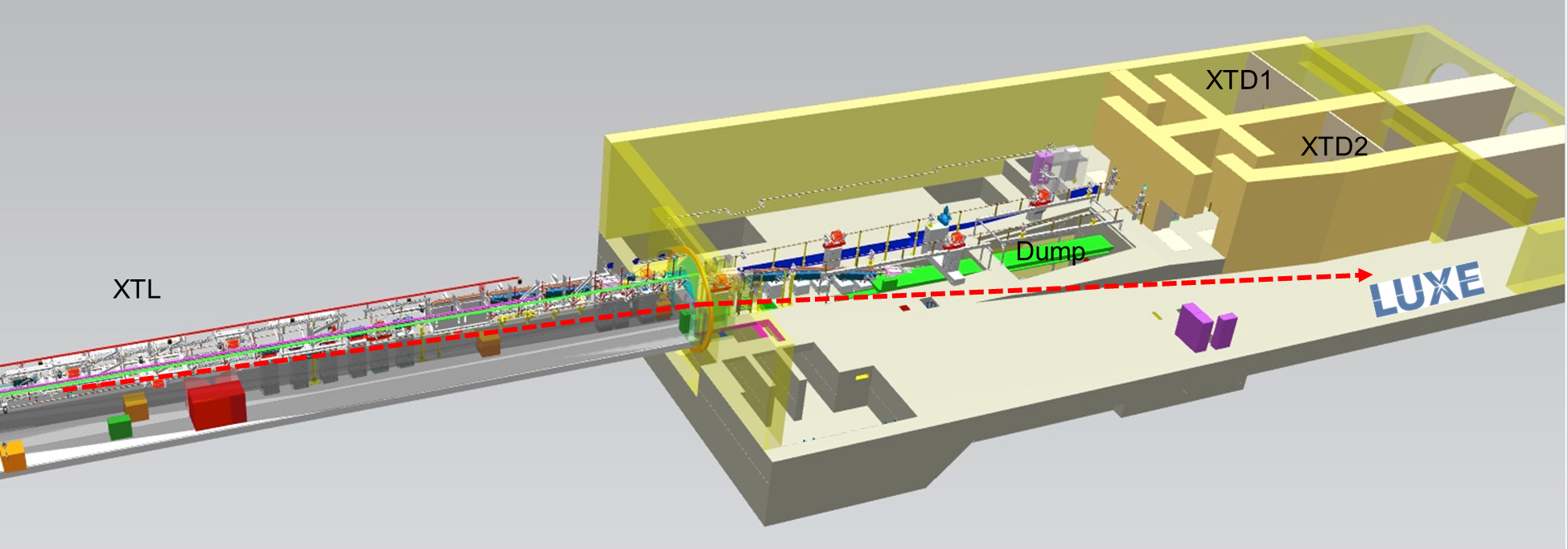}
   \caption{CAD model of the end of the EuXFEL accelerator tunnel and the shaft building with the two existing
beamlines XTD1 and XTD2 to the undulators (SASE1 and SASE2/SASE3) and the XS1 annex, where the LUXE experiment can be installed. The beam extraction and the beam line towards the experiment is sketched with the dashed line.}
   \label{fig:XFEL_sketch_exs}
\end{figure}

The electron bunch for LUXE is extracted with a fast kicker magnet at the end of the Linac tunnel, where the rest of the bunch train proceeds to the fan of undulators via the XTD1 and XTD2 tunnels (see Figure~\ref{fig:XFEL_sketch_exs}).  The LUXE bunch is deflected towards the XS1 building, where the LUXE experiment can be housed. 

As the primary mission of the EuXFEL is the production of x-rays for scientific experiments, the schedule of LUXE installation and later of LUXE experiments must be adapted to its time schedule. Every year there are in excess of $4000$ hours of x-ray delivery, during which also LUXE data taking is possible, making this facility unique when compared to other planned or ongoing efforts~\cite{reis_e-320_2024,Pentia:2023jqf,Mirzaie:2024iey}. As far as access to the tunnel for installation is concerned, there is normally a two-week access each year in the summer and an access of 4-6 weeks in the winter.

The lack of complete control over the accesses which are needed for installation necessitates a flexible attitude, adequate space and equipment to do pre-assembly work and to store pre-assembled components while waiting for the available time windows for tunnel installation.

\section*{Laser and diagnostics} 

A titanium sapphire laser system will be positioned in a ground-level building above the beam extraction tunnel. The initial design foresees this system to operate at 40~TW power ("phase-0"), later upgrading to 350~TW ("phase-1"), with a wavelength of 800~nm and pulse lengths between 25-30~fs. While normally operating with linear polarisation, circular polarisation is preferred for data analysis. This preference and its implications on experiment outcomes, such as pair yields in different setups, are discussed further in Ref.~\cite{Abramowicz:2021zja}.
\begin{figure}[h!]
    \centering
    \includegraphics*[width=0.85\textwidth]{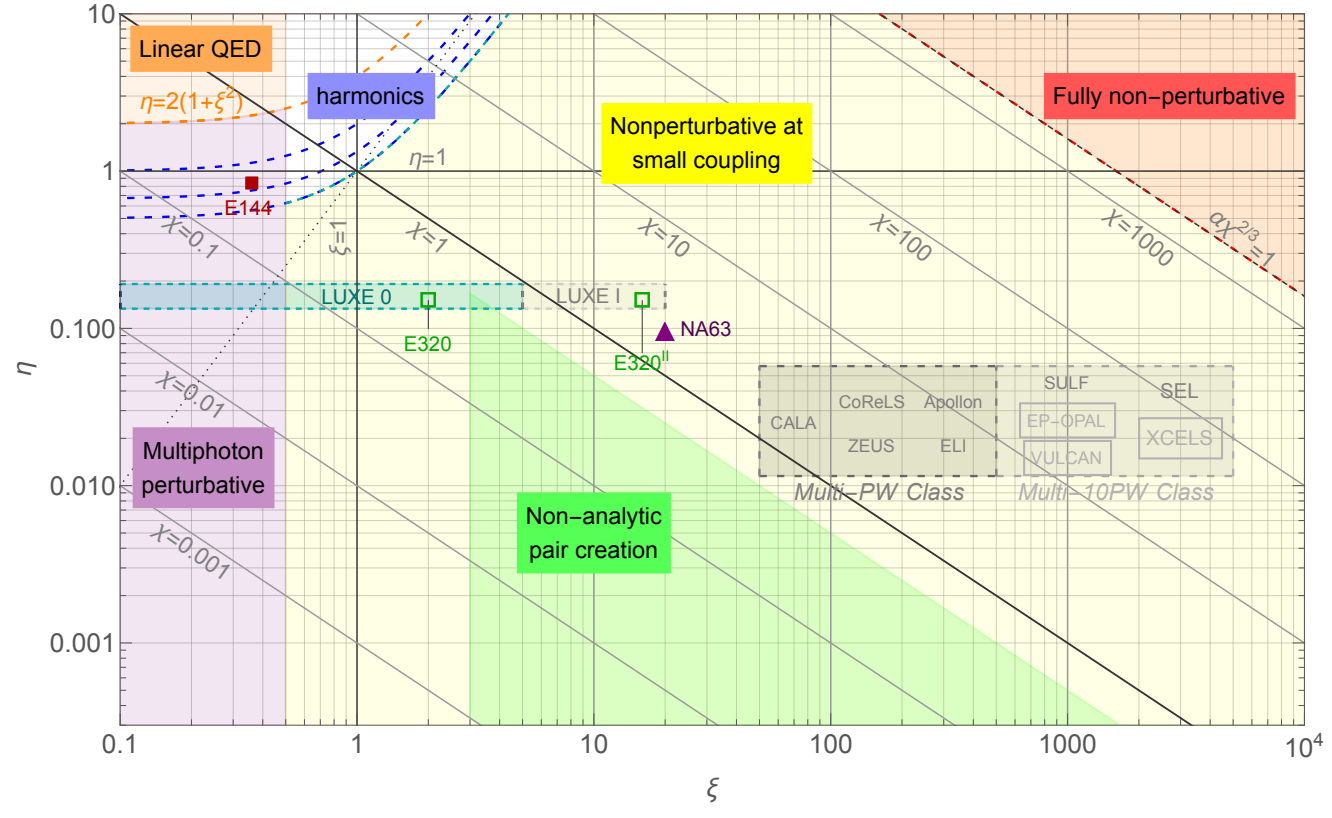}
    \caption{Energy parameter $\eta$ versus the laser intensity parameter $\xi$. The shaded regions marked as ``LUXE~0'' and ``LUXE~I'' show the values accessible at the LUXE experiment for the energy range accessible at the EuXFEL accelerator and possible combinations of laser power and spot size.  Also shown are previous, present and planned experiments~\cite{Bula:1996st,Burke:1997ew,Cole:2017zca,Poder:2018ifi,Chen:22,homma2016combined}. The coloured shaded regions represent different regions of the Breit-Wheeler QED predictions. Adapted from~\cite{Fedotov:2022ely}. }
    \label{fig:megaplot_exs}
\end{figure}

The laser beam is directed to the interaction chamber via a 40~m long vacuum pipe, focusing tightly within a diameter range of $3$ to $150$~$\mu$m in the vacuum chamber, depending on the targeted laser intensity. In the chamber, it intersects the electron beam at a $17.2^{\circ}$ angle, at a 1~Hz repetition rate. Figure~\ref{fig:megaplot_exs} shows the $\xi$ and $\chi$ values that can be accessed at LUXE at the various electron beam energies, laser power and spot size.

The laser will be synchronised with the electron beam via a system  ~\cite{Synchroxfel,Schulz} developed by DESY for the synchronisation of lasers used by the so-called ``pump and probe'' experiments at the EuXFEL. This system has operated for years and demostrated the stable synchronisation of two RF signals to better than 13~fs, compared to a LUXE requirement of 25~fs given the electron and laser bunch lengths and their relative collision angle of $17.2^\circ$.

To ensure LUXE's high-precision aims, a state-of-the-art laser intensity diagnostic system is implemented to monitor shot-to-shot peak intensity fluctuations and longer-term stability. This monitoring is crucial, as the laser intensity is expected to dominate the uncertainties in the measurement of the pair-production rates.  Rather than controlling laser output, the diagnostic system, including optical spectrometers, energy and fluence monitors, and sophisticated \textit{Insight} spatio-temporal reconstruction, will tag each laser shot intensity post-interaction, ensuring a precision of laser peak intensity better than 1\% to reach LUXE's target of 2.5\% relative shot-to-shot precision. A laser pointing stabilisation system using neural networks is also being designed.

\section*{Detectors}

The LUXE experiment is designed to analyse electron-laser and photon-laser interactions as well as photons interacting in the photon beam dump. It is therefore necessary to measure the multiplicities and the energy spectra of electrons, photons and positrons. The fluxes of these particles vary strongly depending on the running mode and the location. A full simulation of the experimental area was performed with \geant to decide on the technologies for the detectors. 

\begin{figure}[htbp]
\centering
\includegraphics[width=0.48\textwidth]{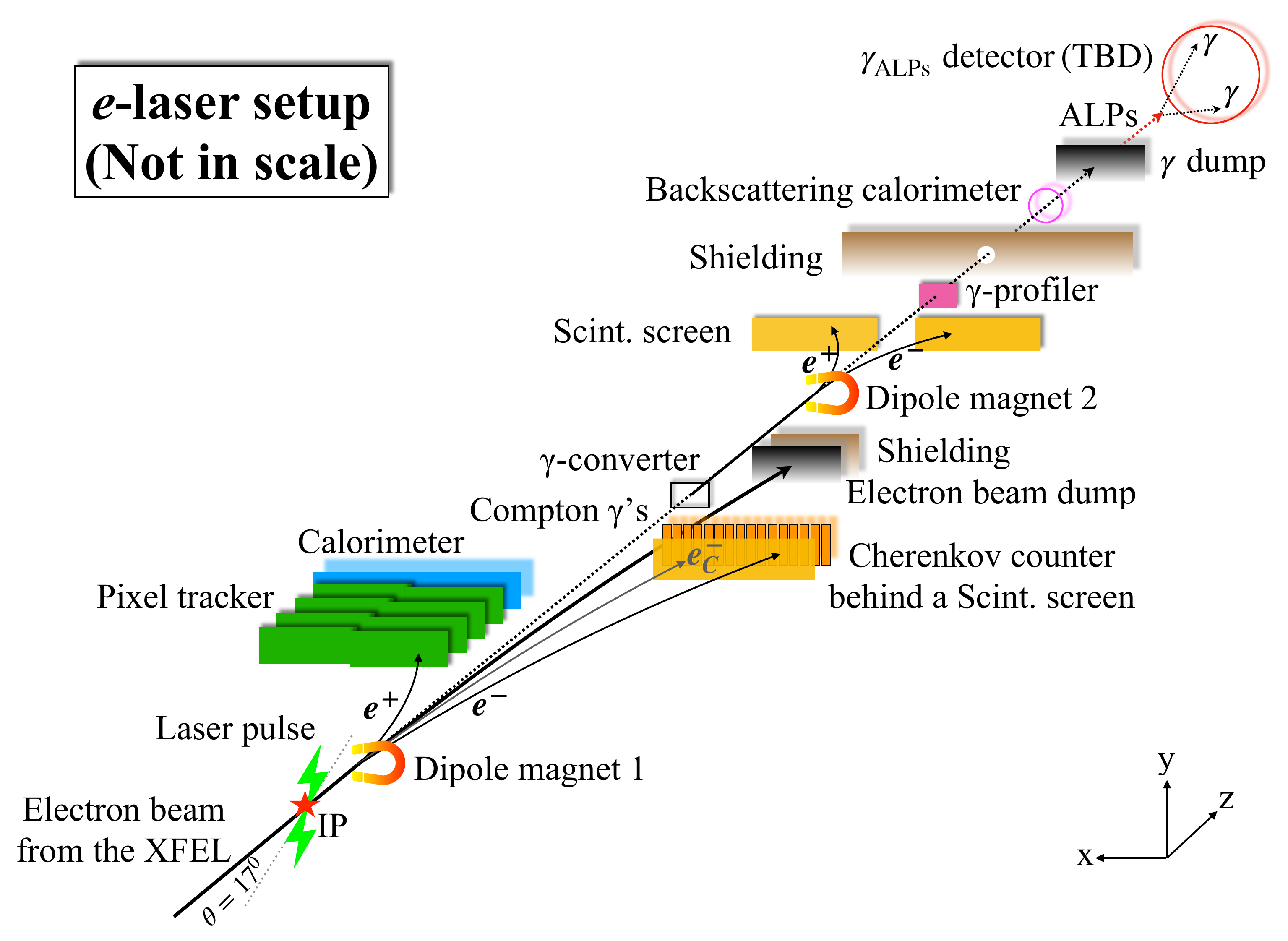} 
\includegraphics[width=0.48\textwidth]{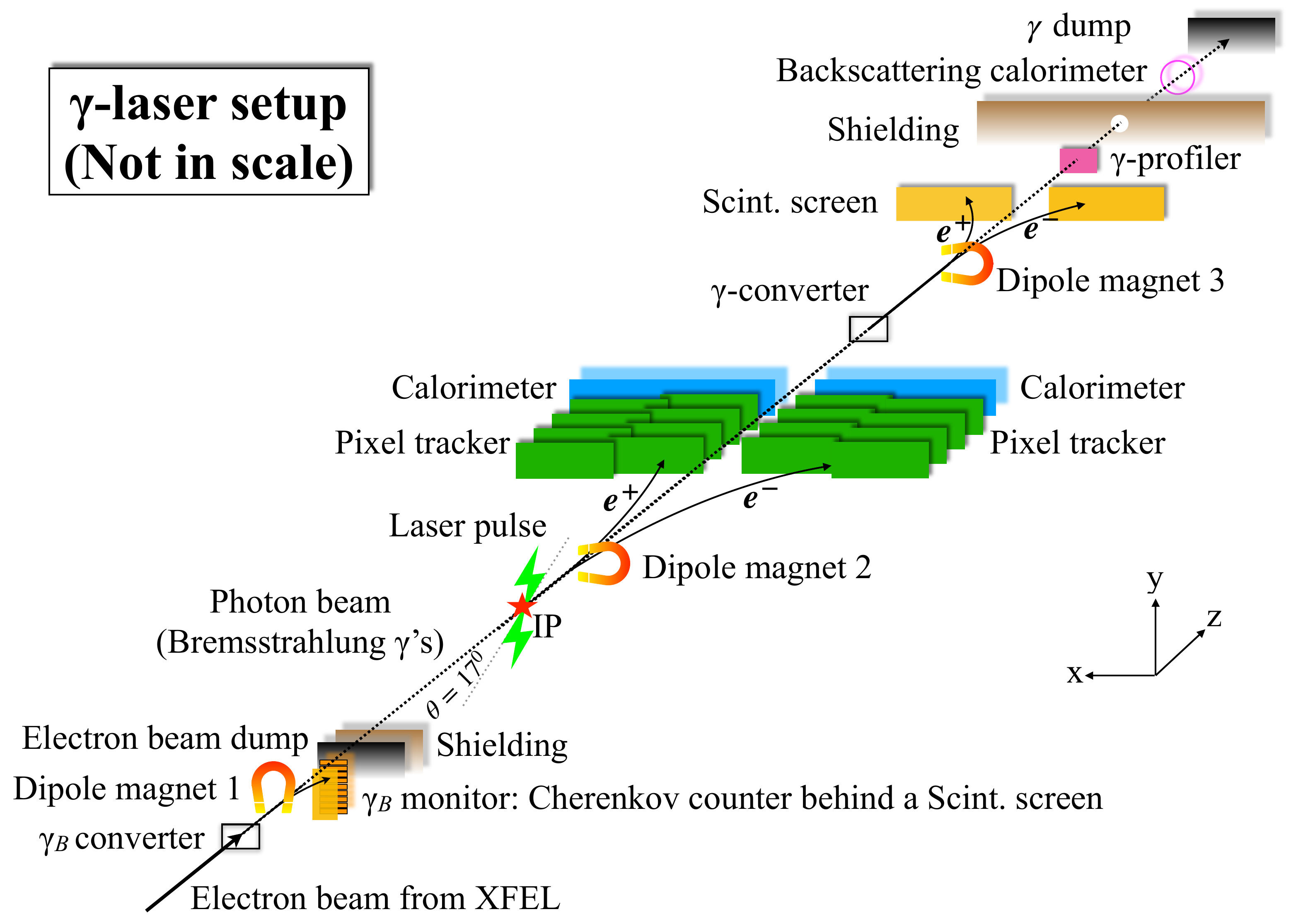} 
\caption{Schematic layouts for the \elaser{} and \glaser{} setup. Shown are the magnets, detectors and main shielding and absorbing elements.}
\label{fig:layoutdet_exs} 
\end{figure}

Sketches of the layout of the experiment for the \elaser and the \glaser modes are shown in Figure~\ref{fig:layoutdet_exs}. 
There are three separate areas in the LUXE experiment that are required for the strong-field QED aspects of the research programme: the target area, the interaction point (IP) area and the photon area.

The \textit{target area} is only relevant for the \glaser{ } mode. A target chamber is installed for the purpose of converting the electron beam to a photon beam either via bremsstrahlung on a tungsten target or via inverse Compton scattering on a low-intensity laser beam. Behind the target chamber is a dipole magnet to separate electrons and positrons from the photon beam. In the \elaser mode the magnet is turned off.
    
The \textit{IP area} hosts the interaction chamber, where the incoming electron or $\gamma$ beam will collide with the high-power laser, and a dipole magnet that separates electrons, positrons and photons. Behind the magnet on the left-hand side ($x>0$, the ``positron side''), there is a tracker and a calorimeter. On the right-hand side ($x<0$, the ``electron side''), the setup depends on whether the data is taken in the \elaser or \glaser mode, as described below.

The pixel tracking detector of LUXE will be built from ALPIDE silicon pixel chips organised into $\sim 27$~cm long and $\sim 1.5$~cm wide ``staves’’. These staves are already being used by the ALICE collaboration in their barrel inner tracking system (ITS) since its upgrade completing in 2021. In LUXE, the staves are organised in a planar (telescope-like) array facing the beam, rather than in a barrel geometry. This required a completely new consideration of the mechanical supporting structure. Furthermore, controlling the pixel clustering and track reconstruction algorithms is much more challenging in LUXE than in ALICE or any other running / near future experiment. That is, due to the signal multiplicity at high $\xi$ values, which is expected to be roughly $\sim\mathcal{O}(50)$ times larger than even the most extreme multiplicities expected in future hadron or muon collider experiments. Specifically, in the densest part of the signal spatial distribution at the detector face, the hit density may be as large as $\sim 100$ particles per mm$^2$. This large signal multiplicity may lead to large ambiguities if not treated carefully. While the clustering used in this work is standard, the track reconstruction algorithm in LUXE uses a Kalman Filter algorithm, which is adapted to deal with this challenge, via the seeding step. The seeding algorithm is completely new and uses the lookup-table concept to deal with the large multiplicity prior to attempting the track fit. Once the track fit is done, quality cuts on track (and fit) parameters are applied to reject background tracks, and combinatorial signal tracks. This helps to achieve track reconstruction efficiency of more than 90\% in the energy range $>2$~GeV, with a strong background rejection (particularly at low particle multiplicities), an energy resolution smaller than 1\% and most importantly, and a linear tracking response to the signal particle multiplicity ranging between 1 and at least 40,000 particles. 
    
In the \glaser case, a tracker and a calorimeter are used on both the positron and the electron side. Due to the low multiplicity of electrons and positrons it is possible to combine the information on the electron and positron and thus reconstruct the energy of the initial photon event by event. 

The design of high-granularity compact electromagnetic calorimeters is driven by the need to identify and measure electromagnetic showers on top of a widely spread low energy background in particular for future electron-positron colliders. This is also true for the LUXE ECALs (called ECAL-P and ECAL-E, respectively for the positron and the electron sides), with the added challenge of having to identify many overlapping showers, up to $10^6$ under certain conditions. The compactness is limited by the inherent Moliere radius of the passive absorber, with tungsten providing the smallest Moliere radius. To prevent the electromagnetic shower from diffusing laterally, the LUXE ECAL-P will be equipped with active sensor planes of thickness less than 1~mm. 
 
The readout will be based on a novel multi-channel, ultra-low power (5\,mW/channel) readout ASIC FLAME, consisting of an analog front-end with variable gain and of a fast sampling (20\,MSps) 10-bit ADC converter in each readout channel. A dedicated version, FLAXE, will be adapted to the LUXE experimental conditions. FLAXE will be among the first few front-end technologies designed for use at future colliders to be applied in a running experiment.

In the \elaser case, the flux of Compton-scattered electrons after the IP varies between $10^5$ and $10^9$, depending on the laser intensity parameter. In this case the tracking detector and calorimeter are replaced by a high-flux detection system consisting of a scintillator screen imaged by an optical camera, and a spatially segmented gaseous Cherenkov detector which measure the flux of particles as a function of their deflection in the dipole magnetic field. The two systems are complementary in that the scintillator screen provides a high spatial resolution, while the Cherenkov detector allows to go to extreme particle rates and has built-in low-energy background rejection because of the Cherenkov energy threshold. The spatial segmentation of the Cherenkov detector is achieved through an array of air-filled reflective tubes which guide the produced Cherenkov light towards a photodetector at the end of the tube. Air as an active medium reduces the amount of Cherenkov light and increases the Cherenkov threshold. A similar set of high-flux electron detectors is used to diagnose the initial gamma beam spectrum in the LUXE \glaser mode.

In general, the environment for the LUXE post-IP particle detectors provides an ideal experimental test bed for detector technologies developed for future collider applications, in particular for high granularity tracking detectors and calorimeters. A strong link exists between LUXE particle detector developments and the CERN DRD collaborations, which should also be fostered in the future.

The \textit{photon area} hosts a photon detection system which is designed to measure the photon flux, angular and energy spectrum of the photons produced at the IP in \elaser mode, or that fly through in the \glaser mode. The energy spectrum is determined using a gamma spectrometer.  The gamma ray spectrometer (GRS) to be implemented at LUXE was conceptualised in~\cite{Fleck2020}. 
The GRS is based on a conversion target and electron and positron spectrometers using a detection system consisting of scintillating screens and cameras. The current conception for LUXE now includes an improved and original reconstruction algorithm based on Bayesian statistics, allowing for a more reliable determination of the reconstruction error compared to the previous “back-substitution” method. This approach can be extended further in future work using machine learning techniques to improve reconstruction accuracy, reliability, and computational speed.  Additionally, due to the nature of the GRS, it is possible to perform shot-to-shot measurements of the photon spectrum without the need for accumulation; particularly relevant for the energies and fluxes anticipated at LUXE. Accompanied with a real-time implementation of the aforementioned reconstruction algorithm, this presents the opportunity for live analysis and diagnosis of the gamma beam, and hence the dynamics at the electron/photon-laser interaction.
The updated design and refined reconstruction algorithm, as presented in the LUXE technical design report~\cite{LUXE:2023crk}, have also been tested at a recent experiment at the Apollon Laser Facility, France. The results presented in~\cite{PhysRevResearch.5.043046} demonstrated the efficacy of the GRS at the 1~GeV scale using a broadband bremsstrahlung source generated by plasma-accelerated electrons. 
 
The spectrometer is followed by a gamma beam profiler (GBP), designed to accurately measure the transverse profile of high-intensity gamma-ray beams with a spatial resolution of 5~\um using sapphire strip detectors. The angular distribution of the Compton photons provides a complementary measure of the laser intensity. Its innovative design enables detailed analysis of laser-beam interactions by providing real-time, on-shot measurements. The GBP's radiation hardness, capable of enduring several MGy, could mark a significant advancement in experimental physics instrumentation.
Sapphire detectors are already well known in high-energy physics for their robustness and radiation hardness, making them ideal for beam monitoring in challenging environments. They are employed, as intense beam condition monitors in FLASH, EuXFEL, and CMS at the LHC. 

Finally, there is a gamma flux monitor (GFM) that faces a unique challenge: the high-flux photon region requires a novel detection technique to overcome the limitations of existing methods. Direct photon counting becomes impractical with photon fluxes exceeding $10^8$~mm$^{-2}$ per bunch crossing (BX) expected at LUXE. The GFM is designed to address this challenge. It exploits an approach by measuring the energy flow of back-scattered particles from the photon beam dump. For this purpose, electromagnetic calorimeter technologies provide a reliable monitoring detector of the gamma flux for the harsh radiation environment of the LUXE experiment. The GFM is realised using a homogeneous calorimeter built of lead-glass blocks with optical readout.
The GFM can also be used at the beginning of a run to optimise collisions by providing feedback on photon flux levels and their spatial distribution. This functionality is analogous to the role of a luminosity monitor in collider experiments.
By using this novel approach and proven construction techniques, the GFM offers a promising solution for monitoring the high-flux photon region in the LUXE experiment.
The gamma profiler and the gamma spectrometer are also sensitive to the photon flux. 

Three beam dumps are required to dump the electron or photon beams. They are designed to minimise back-scattering and backgrounds in the IP and photon area sections.

The LUXE experiment is very modular: it consists of a series of independent subsystems, with redundant measurements of beam and particle properties for cross-calibration and reduction of systematic uncertainties, which are readily accessible when access to the tunnel is allowed. For all components, it is foreseen that the installation is thoroughly prepared and tested on the surface ahead of time so that the time needed for installation in the tunnel is minimised.

\begin{figure}[!h]
\centering
\includegraphics[width=0.48\textwidth]{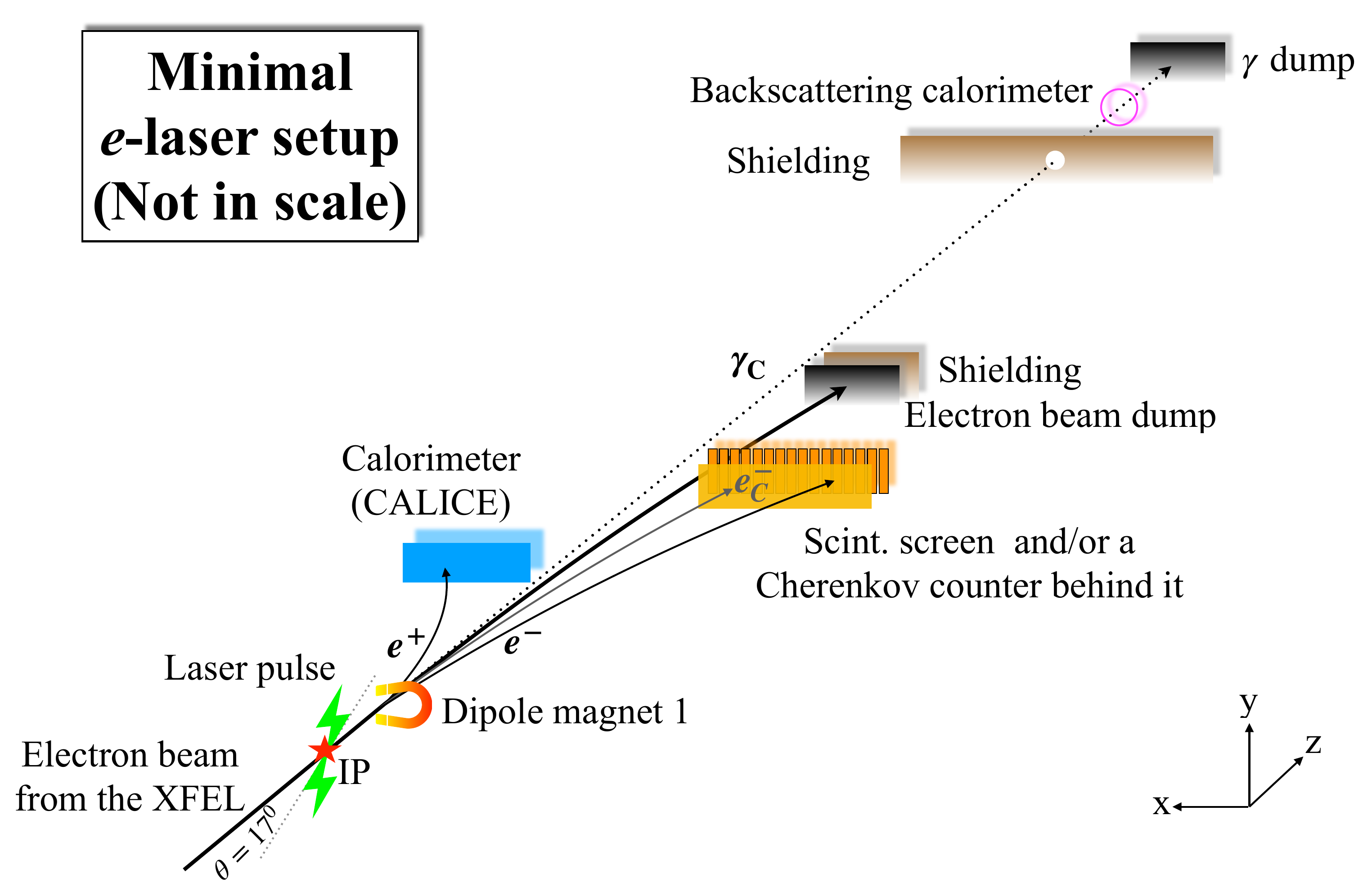} 
\caption{Schematic layout for a minimal version of the LUXE experiment for the \elaser setup. Shown are the magnets, detectors and main shielding and absorbing elements. } 
\label{fig:luxebare_exs} 
\end{figure}

Given the high modularity of the experiment, it is possible to stage the installation, depending on when a given subsystem is required and ready. A \textit{minimal} version of the LUXE experiment was defined as the simplest setup ready for first data. A conceptual sketch of such a minimal version is shown in Figure~\ref{fig:luxebare_exs}.
It is important to stress that such a minimal version is only able to cover a fraction of the LUXE physics programme, but it would already allow some physics measurements (for example Compton events via measurements of the electron spectrum with a scintillation screen and/or Cherenkov detector).

\section*{Readiness, timeline, and expected costs}
The LUXE collaboration has provided conceptual~\cite{Abramowicz:2021zja} and technical~\cite{LUXE:2023crk} design reports outlining the path toward the realisation of the experiment. The installation timeline of LUXE is mainly driven by the operation schedule of the European XFEL. The most critical and time-consuming step is the installation of the electron beam extraction line T20, which requires a 12-week shutdown of the EuXFEL. A minimal LUXE experimental setup after the extraction beamline could be installed in a further 3 to 5 weeks of shutdown. An extended maintenance period of EuXFEL covering the required installation time for T20 and LUXE is subject to approval by the EuXFEL council. If approved, LUXE could be realised at the EuXFEL Osdorfer Born site by 2030. This estimated start time is driven by T20 beamline component lead times, availability of technical personnel and funding acquisition. The T20 extraction beamline is fully funded through the ELBEX grant (Horizon Europe INFRA-2023-DEV-01~\cite{elbex}). A $10\,\mathrm{TW}$ optical laser system (JeTi40\footnote{The last pump laser stage to increase power from $10\,\mathrm{TW}$ to $40\,\mathrm{TW}$ is not moveable from the current location of the system.}~\cite{jena40twlaser}) as well as the baseline detector systems are contributed in-kind by LUXE collaborating institutes.

\begin{figure}[htbp]
\centering
\includegraphics[width=1.15\textwidth]{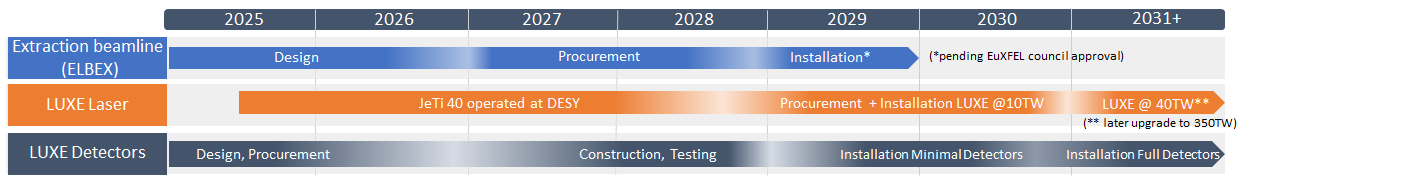} 
\caption{Timeline for the staged construction of the LUXE experiment.} 
\label{fig:luxetimeline} 
\end{figure}

Figure~\ref{fig:luxetimeline} summarises the timeline for the implementation of the LUXE experiment. 

The remaining costs to realise LUXE are considered for a staged scenario:

\begin{itemize}
\item \textbf{Minimal LUXE scenario:} A minimal setup to achieve first collisions between a $10\,\mathrm{TW}$ laser and the EuXFEL beam with minimal set of particle detectors and limited laser intensity diagnostics.
\item \textbf{LUXE Physics scenario:} Experimental setup with $40\,\mathrm{TW}$ laser system, later upgraded to $350\,\mathrm{TW}$, full set of particle detectors and laser diagnostics suite. Aim is to scan the full LUXE strong-field QED parameter space and perform precision measurements. 
\end{itemize}

Table~\ref{tab:luxecost} summarises the cost required for the two LUXE scenarios. 

\begin{table}[h]
    \centering
    \begin{tabular}{|c|c|}
    \hline
    \textbf{Item} & \textbf{Cost [kEUR]} \\\hline\hline
    \textbf{Minimal LUXE scenario} & \textbf{1400} \\\hline
     JeTi40 laser at $10\,\mathrm{TW}$  &  \\ 
     Temporary Laser building &  \\
     Laser beamline &   \\
     Interaction chamber &  \\
     Minimal experiment infrastructure &  \\
     Computing \& DAQ & \\\hline\hline 
     \textbf{LUXE Physics scenario} & \textbf{+1200} \\\hline
     JeTi40 laser at $40\,\mathrm{TW}$  & \\ 
     Precision laser diagnostics &  \\ 
     Experiment infrastructure &  \\\hline\hline 
     \textbf{Upgrade to $350\,\mathrm{TW}$ laser} & \textbf{+5000} \\\hline    
    \end{tabular}
    \caption{Capital costs estimate for different LUXE scenarios.}
    \label{tab:luxecost}
\end{table}

\section*{Summary}
LUXE will probe QED in a new regime of strong fields, by studying collisions between the EuXFEL electron beam, or a high-energy secondary gamma photon beam, with a high intensity optical laser. Running LUXE as a collision experiment in continuous data-taking mode will enable precision measurements of strong-field QED processes, such as nonlinear Compton scattering and Breit-Wheeler pair production. Such investigations will also provide important insights into the modelling of beam-beam interactions at future lepton colliders.

The laser system and particle detectors in LUXE are custom-designed in order to meet the physics goals. LUXE will likely be the first experiment to take precision measurements in a regime of QED never before explored in clean laboratory conditions, with long-term availability of continuous data-taking time and to study high-intensity laser collisions with real high-energy gamma photons.

\printbibliography


\clearpage

\begin{flushleft}
\hypersetup{urlcolor=black}
{\Large \bf The LUXE Collaboration}
\bigskip

\renewcommand{\thefootnote}{\fnsymbol{footnote}}
\begin{center}
\begin{large}

H. Abramowicz~$^{\textrm{\scriptsize 1}}$,
M.~Almanza Soto~$^{\textrm{\scriptsize 2}}$,
M. Altarelli~$^{\textrm{\scriptsize 3}}$,
R.~A{\ss}mann~$^{\textrm{\scriptsize 4}}$,
A.~Athanassiadis\footnote[2]{Also at HU Hamburg}~$^{\textrm{\scriptsize 4}}$,
G.~Avoni~$^{\textrm{\scriptsize 5}}$,
T.~Behnke~$^{\textrm{\scriptsize 4}}$,
M.~Benettoni~$^{\textrm{\scriptsize 6}}$,
Y.~Benhammou~$^{\textrm{\scriptsize 1}}$,
J.~Bhatt~$^{\textrm{\scriptsize 7}}$
% ...

\bigskip

$^{\textrm{\scriptsize 1}}$\textit{Tel Aviv University, Tel Aviv,  Israel}\\
$^{\textrm{\scriptsize 2}}$\textit{IFIC, Universitat de València and CSIC,  Paterna, Spain}\\
%
$^{\textrm{\scriptsize 3}}$\textit{Max Planck Institute for Structure and Dynamics of Matter,  Hamburg, Germany}\\
$^{\textrm{\scriptsize 4}}$\textit{Deutsches Elektronen-Synchrotron DESY, Hamburg, Germany}\\
$^{\textrm{\scriptsize 5}}$\textit{INFN and University of Bologna, Bologna, Italy}\\
$^{\textrm{\scriptsize 6}}$\textit{INFN and University of Padova, Padova, Italy}\\
$^{\textrm{\scriptsize 7}}$\textit{University College London, London,  UK}\\
% ...





\end{large}
\end{center}

\renewcommand{\thefootnote}{\arabic{footnote}}

\end{flushleft}
\end{document}